\documentclass[11pt,twocolumn]{article}
\usepackage[utf8]{inputenc}
\usepackage[english]{babel}
\usepackage[numbers,sort&compress]{natbib}
\usepackage[T1]{fontenc}
\usepackage[top=2cm, bottom=2cm, left=2cm,right=2cm]{geometry}
\usepackage{graphicx}%
\usepackage{xurl}
\usepackage[ruled]{algorithm2e}
\usepackage{algpseudocode}
\graphicspath{{./figures/}}
\usepackage{multirow}%
\usepackage{amsmath,amssymb,amsfonts}%
\usepackage{amsthm}%
\usepackage{mathrsfs}%
\usepackage[title]{appendix}%
\usepackage{xcolor}%
\usepackage{textcomp}%
\usepackage{manyfoot}%
\usepackage{booktabs}%
\usepackage{listings}%
\usepackage{tikz,lipsum,lmodern}
\usepackage{amsmath,amssymb,amsthm}

\theoremstyle{definition}

\usepackage{enumerate}
\usepackage{qcircuit}

\usepackage{caption}
\usepackage{subcaption}
\usepackage[most]{tcolorbox}

\newcommand{\ket}[1]{\ensuremath{\left|#1\right\rangle}} 
\newcommand{\bra}[1]{\ensuremath{\left\langle#1\right|}} 
\newcommand{\braket}[2]{\ensuremath{\left\langle#1|#2\right\rangle}} 

\renewcommand{\bf}[1]{\ensuremath{\mathbf{#1}}}

\providecommand{\keywords}[1]
{
  \small	
  \textbf{\textit{Keywords---}} #1
}
\title{On the explainability of quantum neural networks based on variational quantum circuits}

\author{Ammar Daskin\thanks{Dept. of Computer Engineering, Istanbul Medeniyet University, Istanbul, Turkiye, adaskin25@gmail.com}}

\date{}
\begin{document}

\maketitle

\begin{abstract}
Ridge functions are used to describe and study the lower bound of the approximation done by the neural networks which can be written as a linear combination of activation functions. If the activation functions are also ridge functions, these networks are called explainable neural networks.

In this brief paper, we first show that quantum neural networks which are based on variational quantum circuits can be written as a linear combination of ridge functions by following matrix notations. 
Consequently, we show that the interpretability and explainability of such quantum neural networks can be directly considered and studied as an approximation with the linear combination of ridge functions. 


\end{abstract}

\section{Introduction}
Neural networks have applications in almost every field of science ranging from health to banking.
The ability to interpret the result of a model and explain the learning behavior may be deemed important especially in critical industries such as medicine and health care\cite{tjoa2020survey,linardatos2020explainable,gilpin2018explaining}. 
Limitations of the approximation rates of the classical neural networks can be understood better by using linear combination of ridge functions as an approximation to neural networks.

The power and the limitations of quantum neural networks are yet to be fully understood although for specific parameterized models such as the Bayesian quantum circuits their expressiveness are investigated \cite{du2020expressive}. The expressivity of the quantum machine learning models are also studied on the Boolean cube\cite{herman2023expressivity}, through Fock state-enhancing\cite{gan2022fock}, by mapping to physical observables\cite{wu2021expressivity}, by using covering number in statistical learning theory\cite{du2022efficient},  and by using Fourier expansions\cite{schuld2021effect}. Their approximation power is investigated in Ref.\cite{goto2021universal}. Note that the expressive power and trainability of analog models are also studied in Ref.\cite{tangpanitanon2020expressibility}.

In this paper, we show that quantum neural networks can be written as a sum of ridge functions by following simple matrix transformations.
Therefore, the math and methodologies that are used to understand classical neural networks can be used to study quantum ones.

\subsection{Approximation with ridge functions}
For a random variable y, if we have the observations $y_1, \dots, y_n$ at points $x_1, \dots, x_n$, a standard regression model can be described by $y_i = \hat{y}_i +r_i$, where $\hat{y}_i$ defines the dependence of $y_i$ on $x_i$
When $x_i$ are univariate real values, the assumption is that the dependence is smooth. This leads the following estimation for the regression model\cite{Buja1989LinearSmoothers}:
\begin{equation}
    E(y\mid x) = f(x),
\end{equation}
where $f$ is a smoothing function. In linear smoothing, $\bf{\hat{y}} = (\hat{y}_1, \dots, \hat{y}_n)^T$ can be written in the form of matrix vector transformation: $\bf{\hat{y}}= S\bf{y}$, where $S$ is the smoother matrix that does not depend on $\bf{y}$. 
When there are more than one predictors, estimating the regression surface is hard because of the curse of dimensionality (the data sparseness in high dimensions)\cite{Friedman1981tu}. The general approach is to use the one-dimensional smoother as the building block in an additive model \cite{Buja1989LinearSmoothers}. Given predictors $x_{ij}$s for each $y_i$ outcome, i.e. $\{y_i, x_{i1}, \dots, x_{ip}\}$, the additive model can be described as:
\begin{equation}
    E(y_i|x_{i1},\dots, x_{ip}) = \alpha+\sum_{j=1}^p f_j(x_{ij})+\epsilon.
\end{equation}
where $\epsilon$ is the inherent error, $\alpha$ is a constant parameter, and $fj$s represent unspecified smooth functions. Fitting can be done by using the backfitting algorithm \cite{Friedman1981tu,Breiman1985Backfitting}
which is in matrix form equivalent to the Gauss-Seidel method in numerical linear algebra\cite{golub2013matrix}.

Generalizing this model leads to the projection pursuit model proposed in \citep{Friedman1981tu} where the regression surface is predicted by a linear combination of ridge functions as in the following form:
\begin{equation}
   f(\bf{x}) =  \sum_{k=1}^K f_k (\bf{w_k}\cdot \bf{x}).
\end{equation}
Here, $\bf{w_k}$s represent weight vectors (projection indices) and $f_k$s are ridge functions \cite{pinkus_2015ridgefunctions, candes1998ridgelets, maiorov1999best, petrushev1998approximation}: 
Any multivariate function  $f_k: \mathcal{R}^d \to \mathcal{R}$.  The vector $w_k$ is called the direction and  $f_k$ gives a constant on certain hyper-planes whose normal vectors are parallel to this direction.
Ridge functions are used in approximation theory, partial differential equations, and neural networks. For instance, a feed forward neural network can be defined as \cite{pinkus_2015ridgefunctions}:
\begin{equation}
    \sum_k \gamma_k \sigma \left( \bf{w_{k}}\cdot \bf{x} + b_k\right),
\end{equation}
where $b_k$, $\alpha_k$, and $\bf{w_k}$ represent parameters that describe the neural network.
$\sigma$ is a univariate function (activation function). The degree of approximation  by $\sigma$ functions can be bounded by the degree of approximation by ridge functions (we refer the reader to Ref.\cite{pinkus_2015ridgefunctions} for the properties and other uses of the ridge functions). 
The lower bound of the approximation of the neural networks can be also studied through the relations of the activation functions with ridge functions (e.g., \cite{maiorov1999lower, ismailov2012approximation, klusowski2017minimax}).

If $\sigma$ is chosen as a ridge function these networks are recently called explainable neural networks \cite{vaughan2018explainable}: e.g. an example architecture which has three important structures, a projection layer, sub-network, and a combination layer is described to learn the following: 
\begin{equation}
    f(\bf{x}) =  \mu + \sum_{k=1}^K \gamma_k f_k (\bf{w_k}\cdot \bf{x}).
\end{equation}
Here, $\mu$ and $\gamma_k$s are shift and scaling parameters, respectively. 
In comparison to standard neural networks, the learning in this model can be understood by the "explainable" features: linear projections and uni-variate functions (in other words, the mechanisms used to learn the model can be clearly explained by studying the constitutions that are ridge functions.). 


\section{Explainability of quantum neural networks}
Quantum neural networks\cite{schuld2014quest,kwak2021quantum,jia2019quantum,massoli2022leap}  are generally based on variational quantum circuits\cite{kandala2017hardware} and can be described by
\begin{equation}
    \bra{\bf{x}}W(\bf{\theta})\hat{O}W(\bf{\theta}) \ket{\bf{x}},
\end{equation}
where $\hat{O}$ represents the measurement operator, \ket{\bf{x}} is the input vector formatted as a quantum state and $W(\bf{\theta})$ is a unitary matrix generated by the quantum gates with the angle values defined by $\bf{\theta}$.
Here, by abuse of the notation, we can consider $\hat{O}$ as a selector set on the parts of $W(\bf{\theta}) \ket{\bf{x}}$: e.g., to obtain the measurement output of the first qubit in \ket{0} state, in vector forms, we select the first half of the output  and combine their squared absolute values.
Then, the output quantum state of the quantum circuit applied to \ket{x} can be rewritten as:
\begin{equation}
\label{eq:output}
\sum_{i\in {\hat{O}}} |\braket{\bf{w_i}}{\bf{x}}|^2 = \sum_{i\in {\hat{O}}} f_i(\braket{\bf{w_i}}{\bf{x}}),
\end{equation}
where $\bra{\bf{w_i}}$ represents a row of the unitary matrix $W(\bf{\theta})$.

In variational quantum circuits, generally any change of the vector element of $\bf{\theta}$ may affect multiple rows of $W(\bf{\theta})$. This can affect the studies that try to understand the quantum neural network model.
Therefore, to make any $f_i$ independent from each other, we can use the linear combination of unitary matrices\cite{childs2012hamiltonian, daskin2012universal}.  By following Ref.\cite{daskin2012universal}, we can write the rows of any matrix  $W(\bf{\theta})$ as the first rows of matrices and combine them on a block diagonal matrix:
\begin{equation}
\mathcal{V} = 
 \left(\begin{matrix}
    \left(\begin{matrix}
    \bra{\bf{w_1}}\\
    \bullet\\
    \vdots\\
    \bullet
    \end{matrix}\right)_{N\times N} & &&\\
    &&\ddots &&\\
     & &&   \left(\begin{matrix}
    \bra{\bf{w_N}}\\
    \bullet\\
    \vdots\\
    \bullet
    \end{matrix}\right)_{N\times N} \\
\end{matrix}\right)_{N^2\times N^2} 
\end{equation}
where $N$ is the dimension of  $W(\bf{\theta})$. Using the direct sum, we can write $\mathcal{V}= \bigoplus_i^N V_i$ with $V_i$ representing the unitary matrix for $\bf{w_i}$.
Note that any $N$-dimensional vector can be formed with $O(N)$ quantum gates as the leading row of a unitary matrix by using its Schmidt decomposition. 
Therefore, the construction $\mathcal{V}$ for a generic  $W(\bf{\theta})$ requires at most $O(N^2)$ gates (See Ref.\cite{daskin2012universal} for complexity analysis of the circuit.).

The equivalent quantum state to the output of $W(\bf{\theta}) \ket{\bf{x}}$ can be generated as a part of the outcome of the following transformation:
\begin{equation}
\ket{\psi} = 
 \mathcal{V}
 \left(\begin{matrix}
 \ket{\bf{x}}\\
 \vdots\\
\ket{\bf{x}}\end{matrix}\right)_{N^2\times1} . 
\end{equation}
In a simplified form let $\ket{\psi_i}$ represents the $i$th element of \ket{\psi} with $0\geq i <N^2$, we can define a new selector operator that selects every $\ket{\psi_i}$, where $i\mod N = 0$. That means we can still use the definition similar to Eq.\eqref{eq:output}: 
\begin{equation}
\label{eq:output2}
    \sum_{ i, i\text{ mod } N = 0}^{N^2-1} |\braket{\bf{w_i}}{\bf{x}}|^2
\end{equation}
Note that by writing a quantum operator in this way, we simply make the weight vectors independent from each other.  That means the impact of an angle-change in one quantum gate can be limited to affect only one vector. Therefore, this model is at least as powerful as using a single unitary matrix where a change may affect multiple rows. 
In addition, studying the approximations in this model may be easier since we can easily see how many independent weight vectors are needed and how each weight vector affects the result and training.

\subsection{Explainability of networks that are represented by exponentials}
Ridge functions are also used to approximate the integral forms of functions\cite{pinkus_2015chapter}: One of the most commonly used one is given by the following Fourier form:

\begin{equation}
    \Psi_{\bf{w_k}} := e^{i\bf{x}\cdot \bf{w_k}},
\end{equation}
Here, the complex plane can be rewritten using the only real valued functions, therefore  $\Psi_{\bf{w_k}}$ can be considered as a ridge function for each $\bf{w_k}$. 
Therefore, quantum neural networks such as \cite{daskin2018simple} can be also explained as an approximation through the linear combination of ridge functions.

This formulation is used in Ref.\cite{schuld2021effect} as a main tool to study the expressive power of data encoding models in variational quantum circuits. 
This can be also used to understand quantum circuits that are defined through the Ising type Hamiltonian\cite{xia2017electronic} or machine learning tasks that are solved through the adiabatic quantum computation \cite{farhi2000quantum,albash2018adiabatic}

\section{Discussion and Conclusion}

For a given set of weight vectors, in literature there are many works on the conditions to approximate a given function with unknown ridge functions. 
For instance\cite{ismailov2007representation}, consider $C(X)$ is the space of continuous functions on X, then for a given $\{\bf{w_1}, \bf{ w_2}\}$, for an appropriate choice of some continuous functions $h_1$ and $h_2$, one can write $g_1(h_1(x)) + g_2(h_2(x))=C(X)$ if and only if the lengths of $h_1-h_2$ paths in $X$ are bounded by some positive integer. 
From the Kolmogorov-Arnold representation theorem \cite{schmidt2021kolmogorov}, we know that any multivariate continuous function can be approximated as a sum of univariate functions.
By this theorem, it can be also explained how the hidden layers in the classical neural networks  help in approximations\cite{schmidt2021kolmogorov, hecht1987kolmogorov,kuurkova1992kolmogorov}.
However, a continuous function may not be represented {exactly} by an approximation with {ridge functions} \cite{schmidt2021kolmogorov,maiorov1999best}.
Therefore, the approximation rates in quantum neural networks that reduce to Eq.\eqref{eq:output} and \eqref{eq:output2}  can be well understood. On the other hand, this indicates that the limitations of the approximations with the ridge functions may also be limitations for these types of quantum neural networks.  

In this brief paper, we show that quantum neural networks can be understood as a linear combination of ridge functions, which is used to understand the interpretability and explainability of the classical neural networks. 
In particular, Eq.\eqref{eq:output} and Eq.\eqref{eq:output2} can be used to describe quantum neural network models. 
In addition, since it can be written in the form of a linear combination of ridge functions, the approximation errors and upper and lower bounds on the errors of quantum neural networks can be studied through this formulation. 

In quantum neural networks which can be reduced to Eq.\eqref{eq:output} and \eqref{eq:output2}, the approximation is done by using $N$ ridge functions. Here, $N$ is in general exponential in the number of qubits and can be considered a very large number. For general function approximations, using these equations may be helpful to make a decision on the size of $N$ and the required number of operations and qubits before any application. 
However, the problems solved by the neural networks are in general not easy to define by functions, therefore it is not easy to decide how many functions (and quantum operations and qubits) are required to make the model more trainable.

\bibliographystyle{ieeetr}
\bibliography{ref}
\end{document}